\documentclass[onecolumn]{aastex62}

\usepackage[section]{placeins}
\usepackage{epsfig}
\usepackage{epstopdf}
\usepackage{graphics}
\usepackage{xcolor}

\usepackage{array}
\usepackage{booktabs}
\usepackage{amsmath,amssymb,amsfonts,amsbsy}
\usepackage{mathrsfs}
\usepackage{url}
\usepackage{multirow}

\usepackage{color}
\usepackage{ulem}
\usepackage{enumerate}

\renewcommand{\vec}[1]{\boldsymbol{#1}}

\newcommand{\dif}{\mathrm{d}}

\renewcommand{\figurename}{Fig.}
\renewcommand{\tablename}{Tab.}

\def\aap{Astron.\ Astrophys.\ }

\def\apjl{Astrophys.\ J.\ Lett.\ }
\def\apjs{Astrophys.\ J.\ Supp.\ }

\begin{document}

\title{Multi-messenger observations support cosmic ray interactions surrounding acceleration sources}

\correspondingauthor{Wei Liu, Qiang Yuan}
\email{liuwei@ihep.ac.cn, yuanq@pmo.ac.cn}

\author{Dong-Xu Sun}
\affiliation{Key Laboratory of Dark Matter and Space Astronomy, Purple Mountain Observatory, Chinese Academy of Sciences, Nanjing 210023, China}
\affiliation{School of Astronomy and Space Science, University of Science and Technology of China, Hefei 230026, China}
\affiliation{Key Laboratory of Particle Astrophysics, Institute of High Energy Physics, Chinese Academy of Sciences, Beijing 100049, China}

\author{Pei-Pei Zhang}
\affiliation{Key Laboratory of Dark Matter and Space Astronomy, Purple Mountain Observatory, Chinese Academy of Sciences, Nanjing 210023, China}
\affiliation{School of Astronomy and Space Science, University of Science and Technology of China, Hefei 230026, China}
\affiliation{Key Laboratory of Particle Astrophysics, Institute of High Energy Physics, Chinese Academy of Sciences, Beijing 100049, China}

\author{Yi-Qing Guo}
\affiliation{Key Laboratory of Particle Astrophysics, Institute of High Energy Physics, Chinese Academy of Sciences, Beijing 100049, China}
\affiliation{University of Chinese Academy of Sciences, Beijing 100049, China}

\author{Wei Liu}
\affiliation{Key Laboratory of Particle Astrophysics, Institute of High Energy Physics, Chinese Academy of Sciences, Beijing 100049, China}

\author{Qiang Yuan}
\affiliation{Key Laboratory of Dark Matter and Space Astronomy, Purple Mountain Observatory, Chinese Academy of Sciences, Nanjing 210023, China}
\affiliation{School of Astronomy and Space Science, University of Science and Technology of China, Hefei 230026, China}


\begin{abstract}
The observations of the energy spectra of cosmic-ray have revealed complicated structures. Especially, spectral hardenings in the boron-to-carbon and boron-to-oxygen ratios above $\sim 200$ GV has been revealed by AMS-02 and DAMPE experiments. One scenario to account for the hardenings of secondary-to-primary ratios is the nuclear fragmentation of freshly accelerated particles around sources. In this work, we further study this scenario based on new observations of Galactic diffuse gamma rays by LHAASO and neutrinos by IceCube. We find that the spectra of cosmic ray nuclei, the diffuse ultra-high-energy gamma rays, and the Galactic component of neutrinos can be simultaneously explained, given an average confinement and interaction time of $\sim 0.25$ Myr around sources. These multi-messenger data thus provide evidence of non-negligible grammage of Galactic cosmic rays surrounding sources besides the traditional one during the propagation.
\end{abstract}

\section{Introduction}
\label{sec:intro}
In recent decades, with the fast development of space-borne and ground-based experiments, cosmic ray (CR) measurements have entered an era with high precision. Spectral hardenings above a few hundred GV for various primary nuclei were observed by a number of experiments \citep{2009BRASP..73..564P,2010ApJ...714L..89A,2011Sci...332...69A,2015PhRvL.114q1103A,2015PhRvL.115u1101A,2019PhRvL.122r1102A,2021PhR...894....1A}. Subsequent softenings around an rigidity of $\sim10$ TV was further revealed \citep{2017ApJ...839....5Y,2018JETPL.108....5A,2019SciA....5.3793A,2021PhRvL.126t1102A}. The measurements of secondary-to-primary ratios of nuclei also 
showed spectral hardenings around hundreds of GV rigidity \citep{2021PhR...894....1A,2022SciBu..67.2162D}. All these features indicate that a modification of the conventional model of CR origin and propagation is necessary.

Diffuse $\gamma$-ray emission in a wide energy band can further constrain the propagation and interaction model of CRs. Measurements of diffuse $\gamma$-ray emission from the Galactic plane up to PeV energies by Tibet-AS$\gamma$ \citep{2021PhRvL.126n1101A}, and most recently by LHAASO \citep{2023arXiv230505372C} found that the fluxes are higher than the conventional model predictions. Similar excesses were also found in the inner Galactic plane at GeV energies by Fermi-LAT \citep{2012ApJ...750....3A,2023arXiv230506948Z}. 
Just recently, the IceCube collaboration reported the detection of a Galactic component of neutrinos using 10 years of data \citep{IceCube2023}, which can crucially test the hadronic interactions of CRs in the Milky Way \citep{2023arXiv230617305D,2023arXiv230617285A,2023arXiv230617275F,Shao:2023aoi}.

The production rate of secondary particles is higher than expectation at high energies indicates a larger grammage experienced by high energy particles. Phenomenologically, a break of the energy dependence of the diffusion coefficient can simply lead to such a result \citep{2012ApJ...752...68V,2020ApJS..250...27B}. A natural physical picture is that CRs were confined and interacting in the vicinity of the acceleration sources before injecting into the Galaxy \citep{2009PhRvD..80f3003F,2016PTEP.2016b1E01K,2017PhRvD..96b3006L,2019PhRvD.100f3020Y,2022PhRvD.105b3002Z}.
Since the CR spectra are harder around sources, such interactions become more and more important with the increase of energy, and can thus account for the spectral hardenings of the B/C and B/O ratios \citep{2023FrPhy..1844301M,2023JCAP...02..007Z}. This model is actually a more realistic version of the nested leaky box model which was popular in a long time of CR studies \citep{1973ICRC....1..500C,2016ApJ...827..119C}.

In this work, we further investigate this model with the newly published results of the Galactic diffuse $\gamma$-ray spectra by LHAASO \citep{2023arXiv230505372C} and the neutrino flux by IceCube \citep{IceCube2023}. The re-analysis of the Fermi-LAT data of the same sky regions of LHAASO was also used \citep{2023arXiv230506948Z}, which enables a wide energy coverage of diffuse $\gamma$ rays. We study how this model confronts with these multi-messenger observations, and derive constraints on the interaction time of particles surrounding the sources.
The rest of this paper is organized as follows. In Section 2, we describe the model setting. We present the results in Section 3, and finally conclude in Section 4.

\section{Model Description}
\subsection{Propagation of CRs}
The propagation of charged CRs in the Galaxy is described by the differential equation \citep{1964ocr..book.....G,2007ARNPS..57..285S}
\begin{eqnarray}
\frac{\partial \psi}{\partial t} & = & 
\nabla\cdot(D_{xx}\nabla \psi-{\boldsymbol V_c}\psi)
+\frac{\partial}{\partial p}p^2D_{pp}\frac{\partial}
{\partial p}\frac{1}{p^2}\psi - \frac{\partial}{\partial p}\left[\dot{p}\psi
-\frac{p}{3}(\nabla\cdot{\boldsymbol V_c}\psi)\right]
-\frac{\psi}{\tau_f}-\frac{\psi}{\tau_r}+q({\boldsymbol r},p)
\ , \label{prop}
\end{eqnarray}
where $\psi$ is the particle density per momentum interval, $D_{xx}$ is the spatial diffusion coefficient, $\boldsymbol V_c$ is the convection velocity, $D_{pp}$ is the diffusion coefficient in the momentum space which described the reacceleration by random waves, $\dot{p}$ is the momentum loss rate, $\tau_f$ is the fragmentation time scale, $\tau_r$ is the decay lifetime for radioactive nuclei, and $q({\boldsymbol r},p)$ is the source function. 
In this work, the diffusion re-acceleration model is adopted, and we neglect the convection effect \citep{2019SCPMA..6249511Y,2020JCAP...11..027Y}. The diffusion coefficient in momentum space is related to the spatial diffusion coefficient via $D_{pp}D_{xx} = \dfrac{4p^{2}v_{A}^{2}}{3\delta(4-\delta^{2})(4-\delta)}$, where $v_A$ is the Alfv\'en velocity, and $\delta$ is the slope of rigidity dependence of $D_{xx}$ \citep{1994ApJ...431..705S}. The spatial diffusion coefficient is further assumed to vary with locations in the Milky Way. Specifically, the diffusion is slower in the disk and faster in the halo, as implied by $\gamma$-ray observations of pulsar halos \citep{2017Sci...358..911A, 2021PhRvL.126x1103A}. We describe the detailed form of the spatial diffusion coefficient in Appendix A.

The spatial distribution of the majority of CR sources is assumed to be axisymmetric, which is parameterized as 
\begin{equation}
  f(r, z) = \left(\dfrac{r}{r_\odot} \right)^\alpha \exp \left[-\dfrac{\beta(r-r_\odot)}{r_\odot} \right] \exp \left(-\dfrac{|z|}{z_s} \right) ~,
\label{eq:radial_dis}
\end{equation}
where $r_\odot \equiv 8.5$ kpc is the distance of the solar system to the Galactic center, and $z_{s}=0.2$ kpc is the characteristic thickness of the perpendicular distribution. Parameters $\alpha$ and $\beta$ are taken as $1.69$ and $3.33$ \citep{1996A&AS..120C.437C}. The injection spectrum of nuclei is assumed to be a power-law function of particle rigidity with an exponential cutoff: 
\begin{align}
Q({\cal R}) \propto \left(\dfrac{{\cal R}}{{\cal R}_0}\right)^{-\nu} \exp \left[-\dfrac{\cal R}{{\cal R}_c} \right].
\end{align}
Besides the smooth distribution of the source population, we also employ a local source to reproduce the bump structures of primary nuclei \citep{2019JCAP...10..010L,2020FrPhy..1524601Y}. This local source contributes barely to secondary particles as well as $\gamma$ rays and neutrinos. It is included for completeness in this work. We describe the setting and parameters of the local source in Appendix B.

We also consider the electron component in this work, which contributes to the diffuse $\gamma$-ray calculation. For details of the electron component, one can refer to \citet{2022PhRvD.105b3002Z}.
We work in a cylindrical geometry with radial boundary $r_h=20$ kpc and halo height $\pm L$. The propagation equation is solved with the DRAGON package \citep{2017JCAP...02..015E}. 

\subsection{Secondary particles}
The inelastic collision of CR nuclei and the interstellar medium (ISM) can produce secondary nuclei, photons, and neutrinos. The source function of secondary nuclei can be written as
\begin{eqnarray}
 Q_{j} = \sum_{i>j} (n_{\rm H} \sigma_{i+{\rm H}\rightarrow j} +n_{\rm He} \sigma_{i+{\rm He}\rightarrow j} )v_i \psi_i,
 \end{eqnarray}
in which $v_i$ is the velocity of the parent nuclei species $i$, $n_{\rm H}$ and $n_{\rm He}$ are the number density of ISM hydrogen and helium, $\sigma_{i+{\rm X}\rightarrow j}$ is the production corss section, and $\psi_i$ is the differential density of particle species $i$. For the secondary production during the propagation, $\psi_i$ is the solution of the propagation equation (1). For the secondary production around sources, $\psi_i=q_i\times\tau$, where $\tau$ is the interaction time. 

The emissivity of $\gamma$ rays or neutrinos for the pion decay is  
\begin{equation}
Q_{\gamma,\nu}\; = \;
\sum_{i = \rm p, He} {\displaystyle \int\limits_{E_{\rm th}}^{+ \infty}} \; d E_i \;
v_i \; \left(n_{\, \rm H}
{\displaystyle \frac{d \sigma_{i + {\rm H} \to j}}{d E_j }} +n_{\, \rm He} {\displaystyle \frac{d \sigma_{i + {\rm He} \to j}}{d E_j }} \right)\psi_i.
\label{sec_source}
\end{equation}
The inverse Compton scattering of relativistic electrons off the interstellar radiation field (ISRF) can also produce high-energy $\gamma$ rays, whose emissivity is given by
\begin{equation}
Q_{\gamma}^{\rm ICS}\; = \; c \int \dif \epsilon ~n(\epsilon) \int \dif E_e \psi_e(E_e) F_{\rm KN}(\epsilon, E_e, E) ~,
\end{equation}
where $n(\epsilon)$ is the number density distribution of the background radiation as a function of energy $\epsilon$. The differential Klein-Nishina cross section $F_{\rm KN}(\epsilon, E_e, E)$ is adopted as the following form
\begin{align}
F_{\rm KN}(\epsilon,E_e,E)=\frac{3\sigma_T}{4\gamma^2\epsilon}\left[2q
\ln{q}+(1+2q)(1-q)+\frac{(\Gamma q)^2(1-q)}{2(1+\Gamma q)}\right],
\end{align}
where $\sigma_T$ is the Thomson cross section, $\gamma$ is the Lorentz factor of electron, $\Gamma=4\epsilon\gamma/m_e$, and $q=E/\Gamma(E_e-E)$. On a separate note, when $q<1/4\gamma^2$ or $q>1$, $F_{\rm KN}(\epsilon,E_e,E)=0$. The line-of-sight integral of the emissivity (divided by $4\pi$) gives the flux of specific direction of the sky. Note that, for ultra-high-energy $\gamma$ rays, the pair production in the interstellar radiation field results in suppression of the spectrum, which needs to be properly included \citep{2006A&A...449..641Z,2006ApJ...640L.155M}.

\section{Results}
The diffuse gamma-rays and neutrinos in the Galaxy principally originate from the interactions between CR proton and helium impinging on the nuclei of the ISM. However currently there are still large uncertainties in the measurements of the local CR energy spectra above tens of TeV, which prevents us from the precise evaluation of the corresponding diffuse gamma-rays and neutrinos. Therefore throughout the evaluation of the diffuse gamma-rays and neutrinos, we take into account the uncertainties of the CR measurements.


Before evaluating the diffuse gamma-rays and neutrinos, the CR spatial distribution in the whole Galactic halo has to be calculated by solving the diffusion equation. 
In the case of secondary production at source, the B/C ratio is not determined solely by the transportation parameters. The injection spectra and the confinement time around the sources also affect the B/C ratio. We fit the B/C ratio and primary CR spectra to get the transportation parameters and injection parameters. 

\begin{figure*}[!htb]
\centering
\includegraphics[width=0.6\textwidth]{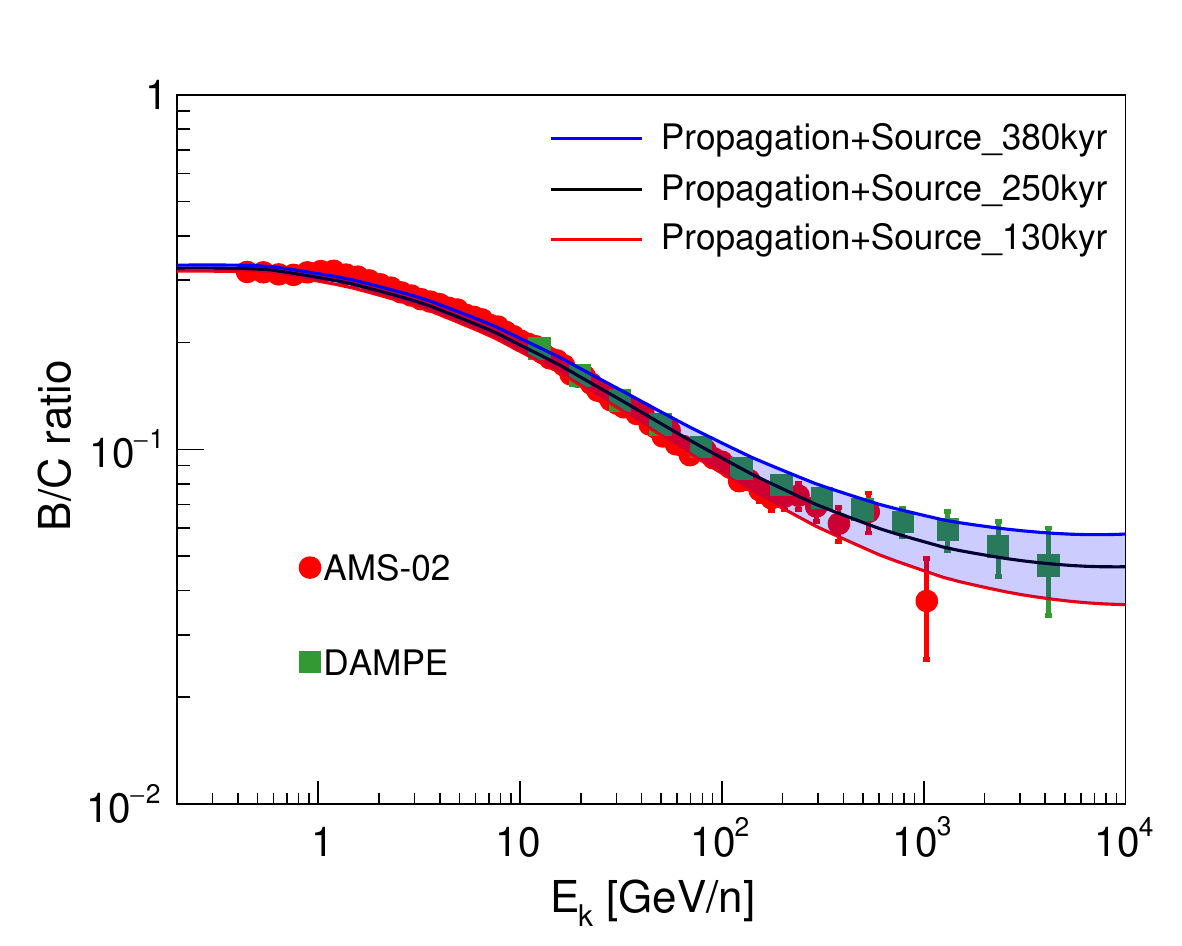}
\caption{The B/C ratio from the model fitting with shaded band representing the envelope of the high and low fittings to the measurements. The data are from AMS-02 \citep{2021PhR...894....1A} and DAMPE \citep{2022SciBu..67.2162D}.}
\label{fig:bcratio}
\end{figure*}

Fig. \ref{fig:bcratio} shows the fitting of the B/C ratio, in which the black solid line is the best fit. The confinement time is estimated to be $\sim 0.25$ Myr. The main uncertainty of this time-scale is believed to be originated from the measurement uncertainties of the B/C ratio. We find that $\tau=0.25\pm0.13$ Myr can reasonably cover the uncertainties of the measurements, as shown by the shaded band.



\begin{figure*}[!htb]
\centering
\includegraphics[width=0.9\textwidth]{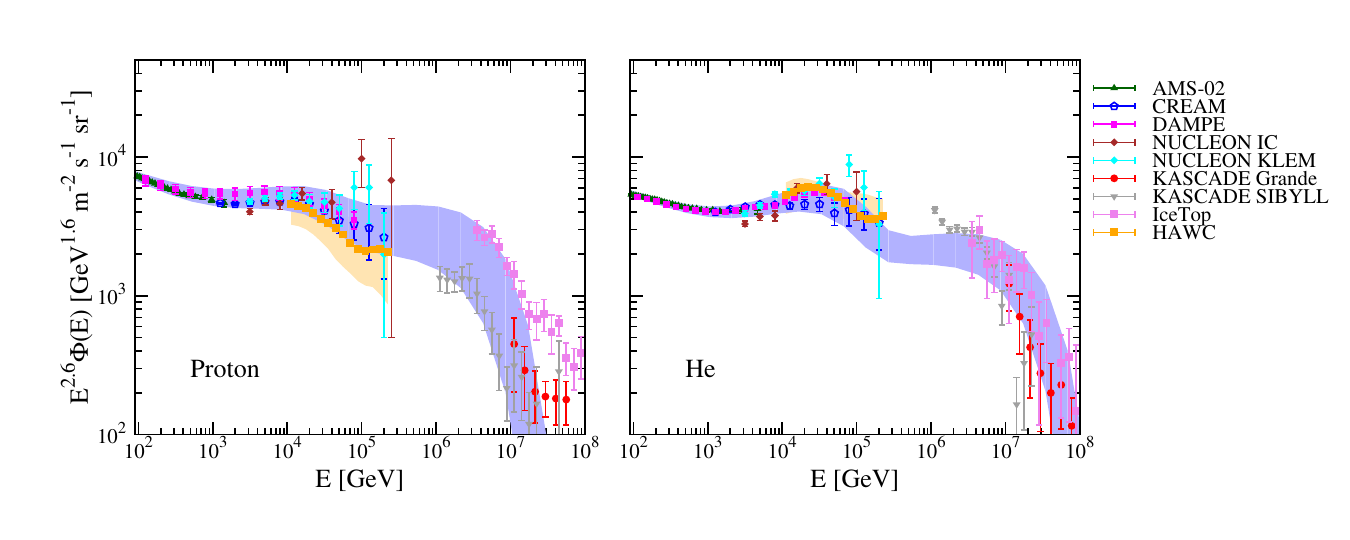}   
\caption{The spectra of protons (left panel) and helium nuclei (right panel). 
Shaded bands show the envelope of the high and low fittings to the data. 
Observational data are from AMS-02 \citep{2015PhRvL.114q1103A,2017PhRvL.119y1101A}, 
DAMPE \citep{2019SciA....5.3793A,2021PhRvL.126t1102A}, CREAM \citep{2017ApJ...839....5Y}, NUCLEON \citep{2019ARep...63...66A}, IceTop \citep{2019PhRvD.100h2002A}, and KASCADE \citep{2005APh....24....1A,2013APh....47...54A}.
}
\label{fig:phe_spect}
\end{figure*}

\begin{figure*}[!htb]
\centering
\includegraphics[width=0.6\textwidth]{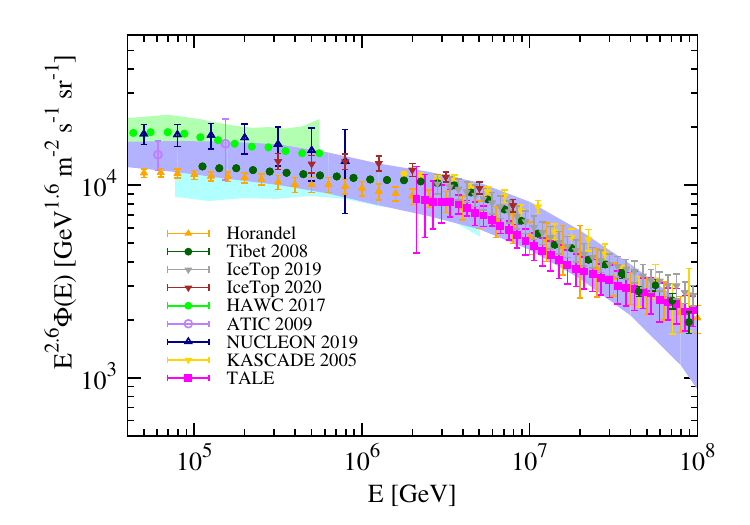}
\caption{The spectra of all particles. The shaded band shows the envelope of the high and low fittings to the data. The data are taken from IceTop \citep{2019PhRvD.100h2002A,2020PhRvD.102l2001A}, Tibet \citep{2008ApJ...678.1165A}, HAWC \citep{2017PhRvD..96l2001A}, ATIC \citep{2009BRASP..73..564P}, NUCLEON \citep{2019ARep...63...66A}, KASCADE \citep{2005APh....24....1A}, TALE \citep{2018ApJ...865...74A}, and a weighted average by \citet{2003APh....19..193H}.
}
\label{fig:all_spect}
\end{figure*}

The fittings of the spectra of protons and helium nuclei are illustrated in Fig. \ref{fig:phe_spect}. The measurements indicates that there are fast falloffs around PeV in both spectra. The position of falloff would significantly affect the yields of the diffuse gamma-rays and neutrinos. But the current measurements are quite inaccurate. To consider this impact on diffuse gamma-rays and neutrinos, the fitting of the energy spectra of both proton and helium are also shown as bands by taking into account the uncertainties of measurements. The high and low values of the injection parameters of the background sources are shown in Table \ref{injec_para}. Here the cutoff rigidity is assumed to be $Z$-dependent. In addition, we also consider the uncertainties of measurements at $\sim$ tens of TeV for different experiments. Therefore the spectral index in the low case is softer than the high one. Furthermore, the results indicate that the cutoff rigidity of the background sources is between $5$ and $7$ PV. To further verify the range of cutoff rigidity, the all-particle spectrum is further calculated. As shown in Fig. \ref{fig:all_spect}, the estimated uncertainties of the all-particle spectrum is compatible with all of the measurements. The injection parameters of the heavy elements are also listed in the Table \ref{injec_para}, all of which are fitted to their separate spectra. The uncertainties of the heavy nuclei are not considered here since they are insensitive to determine the gamma and neutrino fluxes.



\begin{table*}
\centering
\caption{Injection parameters of all major compositions in CRs.}
\begin{tabular}{|c|c c c|c c c|}
\hline
  &  & high & & & low &  \\
\hline
  Element &  ${Q}_0~ [\rm m^{-2}sr^{-1}s^{-1}GeV^{-1}]^\dagger$ & $\nu$  & ${\cal R}_{\rm bc}$ [PV]& ${Q}_0~ [\rm m^{-2}sr^{-1}s^{-1}GeV^{-1}]^\dagger$ & $\nu$  & ${\cal R}_{\rm bc}$ [PV]  \\
\hline
  p    & $4.40\times 10^{-2}$ & $2.45$   &  $7$ &$5.4\times 10^{-2}$ & $2.54$   &   $5$ \\
  He   & $1.96\times 10^{-3}$ & $2.38$  &   $7$ &$5.4\times 10^{-3}$ & $2.42$   &   $5$ \\
  C    & $4.89\times 10^{-5}$ & $2.39$  &   $7$ &$5.4\times 10^{-5}$ & $2.42$   &   $5$ \\
  N    & $3.91\times 10^{-5}$ & $2.46$  &   $7$ &$3.91\times 10^{-5}$ & $2.48$   &   $5$ \\
  O    & $3.91\times 10^{-4}$ & $2.33$  &   $7$ &$3.91\times 10^{-4}$ & $2.47$   &   $5$ \\
  Ne   & $1.96\times 10^{-5}$ & $2.38$  &   $7$ &$1.96\times 10^{-5}$ & $2.44$   &   $5$ \\
  Mg   & $3.42\times 10^{-5}$ & $2.38$  &   $7$ &$3.42\times 10^{-5}$ & $2.42$   &   $5$ \\
  Si   & $3.42\times 10^{-5}$ & $2.49$  &   $7$ &$3.42\times 10^{-5}$ & $2.74$   &   $5$ \\
  Fe   & $1.76\times 10^{-5}$ & $2.25$  &   $7$ &$1.76\times 10^{-5}$ & $2.58$   &   $5$ \\
\hline
\end{tabular}
\label{injec_para}
\end{table*}

\begin{figure*}[!htb]
\centering
\includegraphics[width=0.45\textwidth]{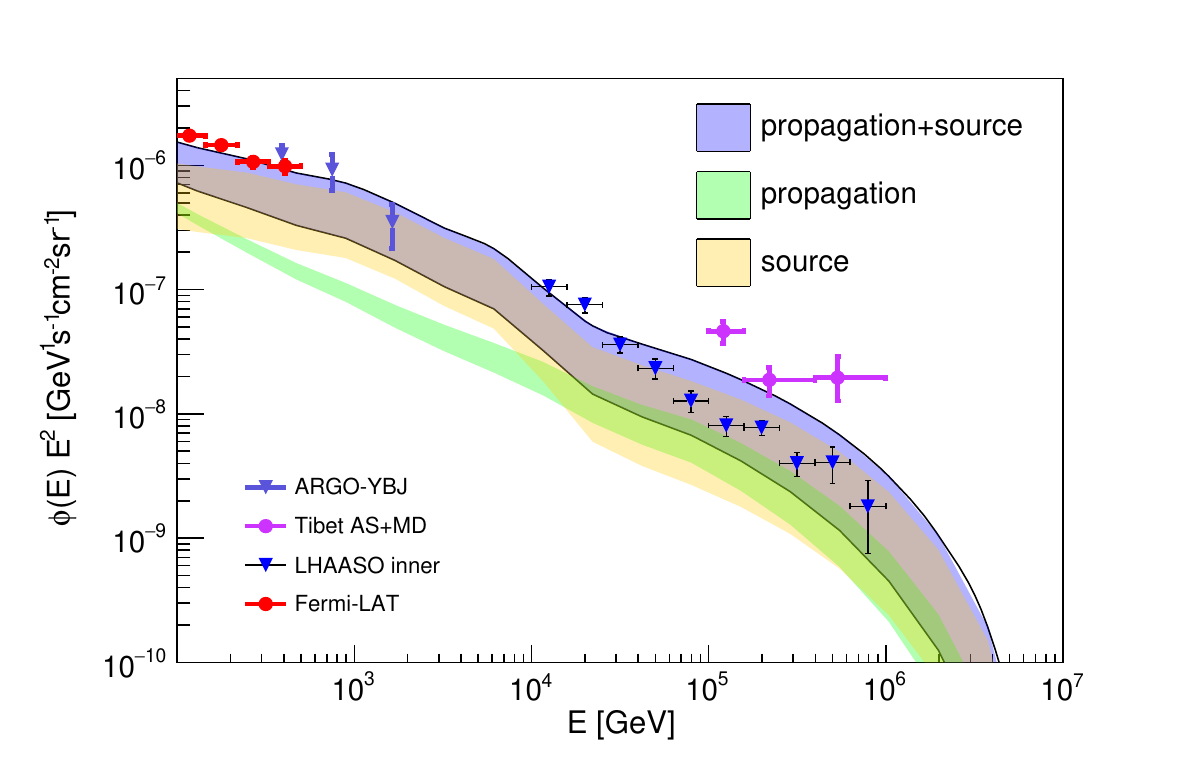}
\includegraphics[width=0.45\textwidth]{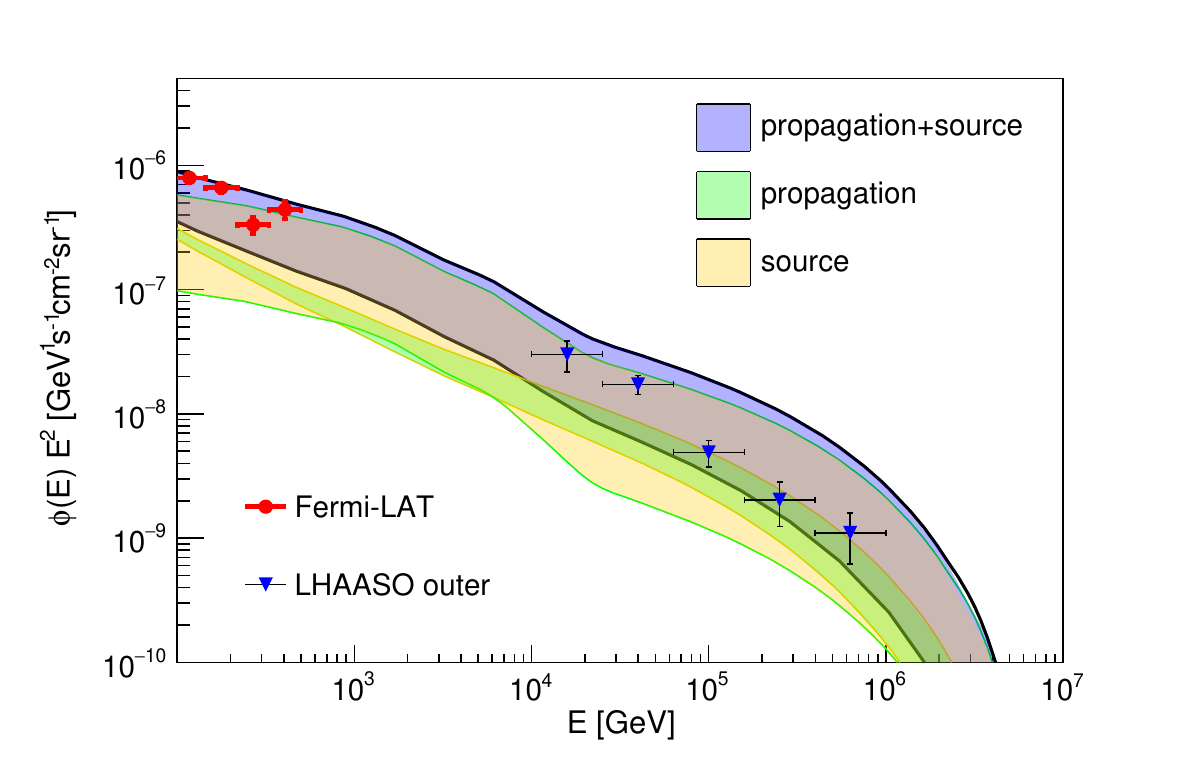}
\caption{The expected diffuse $\gamma$-ray fluxes in the inner (left) and outer (right) Galactic planes, compared with the LHAASO data \citep{2023arXiv230505372C}. The measurements by Tibet-AS$\gamma$ \citep{2021PhRvL.126n1101A} and ARGO-YBJ \citep{2015ApJ...806...20B} for slightly different sky regions are also over-plotted for comparison.
}
\label{fig:dge}
\end{figure*}

In Fig.~\ref{fig:dge}, we demonstrate the expected diffuse gamma rays in the inner ($15^{\circ}<l<125^{\circ}$,
$|b|<5^{\circ}$) and outer ($125^{\circ}<l<235^{\circ}$, $|b|<5^{\circ}$) Galactic planes as defined in \citet{2023arXiv230505372C}. We apply the same masks as adopted in the LHAASO analysis to enable a self-consistent comparison. We find that the diffuse gamma-rays from the CR propagation could not well explain the LHAASO data, except for the extreme case. That indicates the diffuse gamma-rays measured by the LHAASO still contain the contribution from the freshed CRs. Especially at $\sim$ TeV energies, the bumps of the model predictions at $\sim 1$ TeV are from the inverse Compton scattering of high-energy electrons whose spectra show suppression above TeV energies \citep{2017Natur.552...63D}. This is consistent with the breaks measured by the LHAASO.
After taking the contribution from freshed CRs in account, we can see that within the current uncertainty bands which are mainly from the uncertainties of CR spectra, the model can well reproduce the measurements.


\begin{figure*}[!htb]
\centering
\includegraphics[width=0.6\textwidth]{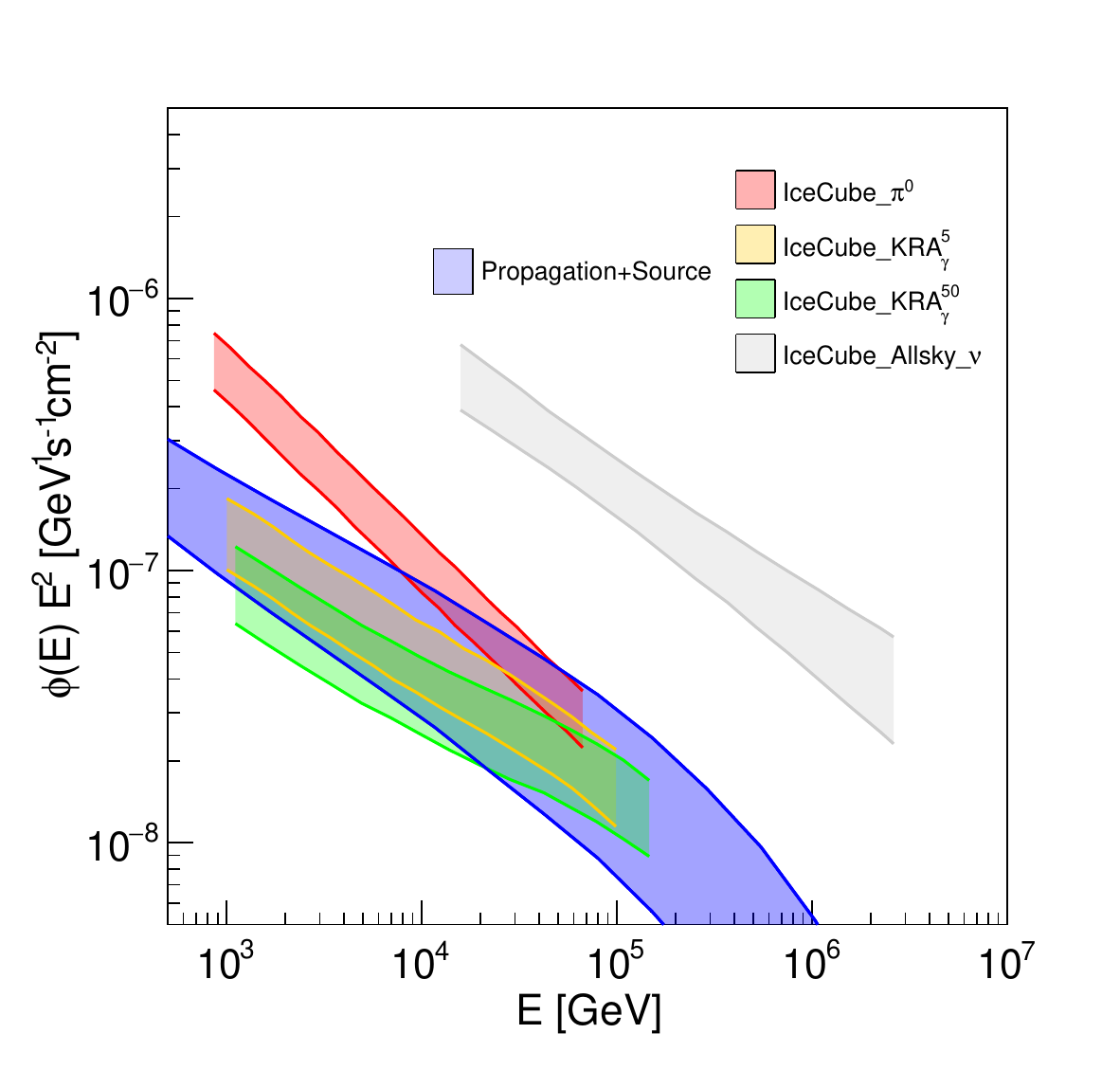}
\caption{The expected neutrino fluxes shown as the blue band, compared with the IceCube measurements for different Galactic model assumptions and the all-sky fluxes \citep{IceCube2023}. 
}
\label{fig:dgn}
\end{figure*}

Within the same model frame we calculate the Galactic neutrino emission. On average, the inelastic $pp$ collision produces nearly one-third neutral pions and two-thirds charged pions. Each neutral pion decays into a pair of gamma rays, whereas each charged pion decays into two muon neutrinos and one electron neutrino (here we do not distinguish neutrinos and anti-neutrinos). We use the \texttt{AAfrag} package \citep{2019CoPhC.24506846K} to calculate the neutrino emission. The neutrino fluxes at production have ratio $\nu_{\mu}:\nu_e\approx 2:1$. After the propagation, the flavor ratio becomes nearly $1:1:1$ for all three kinds of neutrinos. 
The results are shown in Fig. \ref{fig:dgn}. Here we show $1/3$ of the total neutrino fluxes to be compared with the IceCube data \citep{IceCube2023}. Note that the IceCube data depend on model assumptions. Our predicted fluxes are roughly consistent with the measurements \citep{IceCube2023}. 

\section{Conclusion}

More and more measurements with high precision challenge the conventional CR propagation model, such as $O(10^2)$ GV hardenings and $O(10)$ TV softenings of various primary CR nuclei, hardenings of the secondary-to-primary ratios, excess of diffuse $\gamma$ rays, and so on. The model of CR propagation in the Galaxy needs to be refined. In this work, we investigate one of alternative models which includes the secondary production of interactions between freshly-accelerated CRs and surrounding gas. The spectra of CRs around sources are harder than those diffusing in the Galaxy, and can thus be important in contributing secondary particles at high energies. A flat B/C or B/O ratio is expected from the interaction around sources, which, together with the contribution from interactions during the propagation, can explain the hardenings revealed by AMS-02 and DAMPE experiments. The confinement time is estimated to be $\sim 0.25$ Myr via fitting to the data. 

Similar interactions of fresh nuclei with the ISM and/or fresh electrons with the ISRF produce secondary $\gamma$ rays and neutrinos. We find that the model predictions can well reproduce the Galactic diffuse $\gamma$-ray measurements by LHAASO and the neutrino measurements by IceCube. Even in the extreme case with CR spectra as high as the upper bounds of the systematic uncertainties of current measurements, the expected diffuse $\gamma$-ray fluxes from the propagation procedure are still lower than the measurements. Adding the harder component from interactions around sources can match the data well. Particular attention should be paid that, the inverse Compton emission from the fresh electron component gives a bump feature of the diffuse $\gamma$-ray emission around TeV, which is just required by the Fermi and LHAASO data \citep{2023arXiv230506948Z}. All these multi-messenger observations of secondary particles suggest that the interactions of CRs around the sources are important at high energies ($\gtrsim$TeV).




\section*{Acknowledgements}
This work is supported by the National Natural Science Foundation of China (Nos. 12220101003, 
12275279, U2031110) and the Project for Young Scientists in Basic Research of Chinese 
Academy of Sciences (No. YSBR-061).

\bibliographystyle{aasjournal}
\bibliography{refs}

\appendix
\section{Spatially dependent diffusion}
More and more observations show that the diffusion of CRs in the Milky Way should be spatially dependent. The observations of very-high-energy $\gamma$-ray emission from pulsars suggest that particles diffuse significantly slower around pulsars \citep{2017Sci...358..911A, 2021PhRvL.126x1103A} than in the average ISM \citep{2017PhRvD..95h3007Y}.
Thus a two-halo model with a slow diffusion disk and a fast diffusion halo might be a more realistic approach of the propagation \citep{2012ApJ...752L..13T}. This inhomogeneous propagation model was later found to be able to suppress the expected large-scale anisotropy amplitudes at high energies \citep{2019JCAP...10..010L, 2019JCAP...12..007Q}. 
Following previous work \citep{2018PhRvD..97f3008G}, we parameterize the spatially-dependent diffusion coefficient as
\begin{equation}
  D_{xx}(r,z, {\cal R} )= D_{0}F(r,z)\beta^{\eta} \left(\dfrac{\cal R}
  {{\cal R}_{0}} \right)^{\delta_{0}F(r,z)},
  \label{eq:diffusion}
\end{equation}
in which the function $F(r,z)$ is written as
\begin{equation}
  F(r,z) = \left\{
\begin{array}{ll}
  {\dfrac{N_m}{1+f(r,z)}+\left[1-\dfrac{N_{m}}{1+f(r,z)}\right]}\left(\dfrac{{z}}{\xi L} \right)^{n},  &  {{|z|} \leq \xi L} \\
\\
  1,  &  { {|z|} > \xi L} \\
  \end{array}.
  \right.
\end{equation}
The propagation parameters are $D_0=4.87\times10^{28}$ cm$^2$~s$^{-1}$ at ${\cal R}_0=4$ GV, $\delta_0=0.65$, $L=5$ kpc, $N_m=0.57$, $\xi=0.1$, $n=0.4$, $v_A=6$ km~s$^{-1}$.


\section{Local CR source}
A local source was introduced to account for the spectral bumps of protons and helium nuclei \citep{2019JCAP...10..010L,2020FrPhy..1524601Y}. It may also explain the spectral anomalies of electrons and/or positrons assuming that primary electrons were accelerated by the supernova remnant and $e^+e^-$ pairs were accelerated by the associated pulsar wind nebula \citep{2021JCAP...05..012Z}. We take Geminga as an illustration of the local source. The characteristic age is inferred from the spin-down luminosity of the Geminga pulsar, $\tau_{\rm src} \sim 3.4 \times 10^5$ years \citep{2005AJ....129.1993M}, and its distance to the solar system is taken as $330$ pc. The injection of CRs is approximated as burst-like, and the spectrum is assumed to be a power-law function with exponential cutoff
\begin{equation}
q_{\rm inj}({\cal R}, t)=q_{0} \delta(t-\tau_{\rm src}) \left(\dfrac{\cal R}{{\cal R}^{\rm src}_0}\right)^{-\gamma} \exp \left(-\dfrac{\cal R}{{\cal R}^{\rm src}_c}\right)~,
\label{eq:nearby}
\end{equation}
where ${\cal R}^{\rm src}_c$ is the cutoff rigidity. The propagated spectrum from the local source can be calculated using the Green's function method, 
\begin{equation}
\psi(r,{\cal R},t)=\frac{q_{\rm inj}({\cal R})}{(\sqrt{2\pi}\sigma)^3}
\exp\left[-\frac{(\vec{r}-\vec{r_{\rm src}})^2}{2\sigma^2}\right],
\end{equation}
with $\sigma = \sqrt{2D_{xx}({\cal R})(t-\tau_{\rm src})}$. Here we assume a spherical geometry with infinite boundary. The diffusion coefficient is taken as the disk value in the local vicinity. The injection spectral parameters of the local source are given in Table~\ref{injec_para_local}.

\begin{table*}
\centering
\caption{Injection parameters of major compositions for the local source.}
\begin{tabular}{|c|c c c|}
\hline
  Element &  $q_0~[\rm GeV^{-1}]$ & $\gamma$ &  ${\cal R}^{\rm src}_c$~[TV]   \\
\hline
  p    & $1.2\times 10^{52}$ &  $2.22$  & $18$ \\
  He   & $1.2\times 10^{51}$ &  $2.25$  & $18$ \\
  C    & $3.0\times 10^{50}$ &  $1.60$  & $18$ \\
  N    & $8.1\times 10^{49}$ &  $2.10$  & $18$ \\
  O    & $1.5\times 10^{50}$ &  $2.10$  & $18$ \\
  Ne   & $1.1\times 10^{50}$ &  $1.80$  & $18$ \\
  Mg   & $1.1\times 10^{50}$ &  $2.00$  & $18$ \\
  Si   & $1.1\times 10^{50}$ &  $2.10$  & $18$ \\
  Fe   & $1.8\times 10^{49}$ &  $2.00$  & $18$ \\
\hline
\end{tabular}
\label{injec_para_local}
\end{table*}

\end{document}